# Competition between clustering and dispersion of cobalt atoms on perovskite surfaces: SrTiO$_3$(001) and KTaO$_3$(001)

*Aji Alexander[1], Pankaj Kumar Samal[1], Llorenç Albons[1], Jesús Redondo[1,2], Jan Škvára[1], Igor Píš[3,4], Lukáš Fusek[1], Josef Mysliveček[1], Viktor Johánek[1], Dominik Wrana[1,5]\*, Martin Setvín[1]*

[1] Department of Surface and Plasma Science, Charles University, 18000 Prague, Czech Republic

[2] Department of Polymers and Advanced Materials, Facultad de Química, University of the Basque Country UPV/EHU, 20018 San Sebastián, Spain

[3] CNR – Istituto Officina dei Materiali (IOM), 34149 Trieste, Italy

[4] Institute of Electrical Engineering, Slovak Academy of Sciences, 841 04 Bratislava, Slovakia

[5] Marian Smoluchowski Institute of Physics, Jagiellonian University, 30-348 Kraków, Poland

\* dominik.wrana@uj.edu.pl

**Funding:** Czech Science Foundation (GACR 20-21727X); Czech Ministry of Education, Youth and Sports (QM4ST, CZ.02.01.01/00/22_008/0004572); Grant Agency of Charles University (GAUK 326122); Polish National Science Centre (SONATA 2022/47/D/ST5/02439); Elettra Sincrotrone Trieste (IUS P2022003).

Perovskite oxides are attractive for reactions in photo/electrocatalytic schemes, and extrinsic doping is a common strategy for tuning their properties. It is widely known that extrinsic dopants impact the structure and stability of perovskite surfaces, but an atomic-scale view is missing. Here, noncontact atomic force microscopy (ncAFM) and photoelectron spectroscopy (XPS/PES) are used to combine microscopic and spectroscopic evidence of cobalt adsorption, incorporation, and clustering at surfaces of two prototypical perovskites SrTiO$_3$ and KTaO$_3$. A number of different sub-ML coverages and temperatures of annealing were investigated.

Several common features are observed: cobalt shows a strong preference for ionic nature (+2 and +3 charge states), and remains dispersed as single atoms to a certain extent in both perovskites. Two competing mechanisms are observed upon annealing: coalescence into clusters with a mixed metallic/ionic character, and incorporation into the surface and subsurface regions. The latter is more pronounced in SrTiO$_3$, where a cobalt-stabilized surface reconstruction is identified, whereas for KTaO$_3$ cobalt likely incorporates in the near-surface region.

## 1. Introduction

Perovskite oxides (general formula ABO$_3$)[1] are attractive for catalysis,[2,3] photocatalysis[4,5] and electrocatalysis[3] due to their high catalytic activities in a broad range of chemical reactions, tuneable chemical properties, and unique physical properties such as ferroelectricity[6] or a long diffusion length of charge carriers.[7] One of the main impediments to the practical use of perovskites is the inherent instability of their surfaces,[8-11] which often leads to deactivation of the surface catalytic properties, and eventually, degradation of the material. Co-doping of perovskites at both A and B sites has become an established way to counteract this problem.[12,13] It includes strategies such as reducing the bulk lattice strain by tuning the cation sizes,[12] tuning the surface composition to compensate surface polarity,[14-16] and using exsolution to controllably disperse A and B cations on the surface.[13,17,18] The latter process is an intensive subject of investigation since perovskites show the ability to reversibly produce finely dispersed precious metals (such as Pd) at their surface with very low initial material loadings (0.05 at.%)[19]. This was also observed for the case of stabilization of Rh isolated sites on calcinated powders of Rh-doped SrTiO$_3$[20]

Here we investigate the interaction of various phases of cobalt-based co-catalysts with the KTaO$_3$(001) and SrTiO$_3$(001) surfaces to probe the possibility to stabilize the perovskite surface and disperse the co-catalyst at the atomic scale. Cobalt is attractive for the water gas shift reaction or conversions of alcohols[21-23] and heavy hydrocarbons,[24,25] and is *en route* to become a strategy raw material.[26,27]

KTaO$_3$(001) and SrTiO$_3$(001) can be considered as prototypes of polar and weakly-polar surfaces, respectively. Ionic surfaces and interfaces need to fulfil the Tasker polarity criteria[28-30] to achieve thermodynamic stability. The bulk-terminated KTaO$_3$(001) surface carries a net charge of 1 electron per unit cell for the (KO)$^{-1}$ and (TaO$_2$)$^{+1}$ terminations, while the SrTiO$_3$(001) surface is weakly polar,[31] *i.e.*, it only carries a residual fractional charge that originates from the different ionicity of the SrO and TiO$_2$ terminations. Thus, the bulk-terminated KTaO$_3$(001) surface interacts strongly with adsorbed species and finds different mechanisms to maintain a zero net charge,[32-34], whereas the SrTiO$_3$(001) surface is expected to be less reactive. Here we investigate whether uncompensated surface polarity can be compensated by extrinsic doping, using an ion with a suitable valency. Cobalt can adopt oxidation states of 2+ or 3+, or stay metallic, therefore replacement of atoms in the parent matrix by cobalt has a potential to (i) stabilize the surface, (ii) disperse the material used for compensation and prevent its sintering. We use SrTiO$_3$ to benchmark the behaviour of Co on polar KTO against a more chemically stable bulk termination.

A combination of scanning tunnelling microscopy and atomic force microscopy shows that the majority of the as-deposited cobalt is atomically dispersed at room temperature, while annealing above 300°C results in partial clustering. X-ray photoelectron spectroscopy (XPS) shows that the dominant oxidation state of the ionic cobalt is 2+ or 3+, while the metallic state represents a minor species through a wide range of coverages and temperatures. For SrTiO$_3$, a certain level of cobalt incorporation in the surface is observed even after high-temperature annealing, resulting in a new surface reconstruction. This is not observed for KTaO$_3$, where the surface prefers the bulk termination.

Our results shed light on the atomic-scale details of processes occurring on the systems that favour many catalytic reactions, such as oxidative coupling[33] and photo-catalysis.[35,36] Clusters of metals such as cobalt, nickel, and iron[37-39] are often used in the concept of exsolution,[40,41] where the catalytically active metal changes between a metallic state clustered at the surface and an ionic state dissolved in the bulk.

## 2. Experimental setup

Experiments were performed on single-crystal perovskite samples. KTaO$_3$ was grown by solidification from the nonstoichiometric melt,[42] and *n*-type conductivity was achieved by trace doping (<0.2%) of Ca, or Sr. SrTiO$_3$ was purchased from SurfaceNET, *n*-type samples doped with 0.5% Nb were used. Atomically flat and clean surfaces are required to ensure uniform deposition and distributions of supported metal atoms and for effective STM/AFM characterization. The KTaO$_3$(001) surface was prepared by in-situ cleaving,[44] which provides a bulk-terminated surface containing domains of KO and TaO$_2$ terminations.[32] Before cobalt deposition, the surface was exposed to 170 L of water vapor, which results in the formation of a homogeneous hydroxylated termination with a (2×1) periodicity.[32] This termination was used as a starting point in this study because it seems to represent a realistic model of KTO surface structure under ambient conditions.

For the SrTiO$_3$(001) surface, a preparation method based on wet chemical techniques was used, which is a widely used procedure for this substrate.[45-47] The initial single crystal was cut and mechanically polished along the (001) plane, followed by boiling for 120 minutes in deionized water. The sample was subsequently introduced to an ultrahigh vacuum (UHV) chamber and annealed to 800°C in 5.5×10$^{-7}$ mbar of O$_2$ to remove any carbon contamination. The resulting samples were analysed by XPS to exclude any extrinsic contamination, see Figure S1. This preparation results in a mixture of coexisting surface phases such as c(4×2), c(6×2), and (4×4).[48-51] In principle, annealing to higher temperatures (above 900°C) is known to result in stable reconstructions such as √5×√5 and √13×√13,[51-53] but here we have purposely avoided these two polarity-compensated, sub-stoichiometric reconstructions.[54]

Surface morphology was investigated in a double-vessel UHV system equipped with a commercial cryogenic STM/AFM head (Scienta Omicron Polar SPM Lab). The base pressure was 1×10$^{-10}$ mbar. The KTaO$_3$ bulk crystals were first outgassed in UHV at temperatures between 600°C and 650°C and subsequently cleaved at room temperature by the tungsten carbide blade of a UHV mechanical cleaver.[44] Sample annealing was performed in a manipulator equipped with a BN heater; the sample was kept at the specified temperature for 15 minutes in each annealing cycle. The absolute error of the quoted temperatures is estimated as ±20°C. Cobalt was deposited by physical vapour deposition (PVD), sublimating a 1 mm cobalt wire from a custom-design e-beam evaporator. The deposited amount was monitored by quartz crystal microbalance (QCM). One monolayer (ML) is defined as one atom per surface unit cell of the underlying perovskite, *i.e.*, 6.21×10$^{14}$ atoms /cm$^2$.

Tuning-fork-based AFM sensors with a separate wire connection for detection of the tunnelling current were used (k = 1900 N/m, f$_0$ = 32KHz, Q ≃ 20,000),[55] and the deflection signal was measured by a differential cryogenic preamplifier.[56] Electrochemically etched tungsten tips were used, their metallic character was ensured by repeated treatments on the Cu (110) surface. The ncAFM images shown in this article were measured in both constant frequency and constant height modes.

The chemical characterization of the sample was carried out using two laboratory XPS systems. The first system features a non-monochromatic X-ray source with an Al anode (hν = 1486.4 eV), a SPECS Phoibos MDC 9 electron energy analyzer, configured for both normal and grazing detection angles, and a base pressure below 1×10$^{-9}$ mbar. The second system is equipped with a monochromatic X-ray source with Al kα, a hemispherical energy analyzer EA 15 UHV (PREVAC system) for normal detection angles, and a base pressure below $1\times10^{-9}$ mbar. In the case of monochromatic X-ray source, data were obtained with a pass energy of 100 eV. Apart from a wide survey spectrum, spectral regions of Co 2p, K 2p, Sr 3d, Ti 2p, Ta 4f, O 1s, and C 1s were acquired.

To probe the chemical composition of cobalt coverages as low as 0.05 ML, synchrotron radiation-excited photoelectron spectroscopy (SRPES) with a high X-ray flux and tuneable excitation energy was chosen (BACH beamline of CNR at the Elettra synchrotron in Trieste, Italy). The UHV system had a base pressure of $4\times10^{-10}$ mbar and was equipped with a mechanical cleaver and a Scienta R3000 hemispherical analyzer positioned at a 60° angle from the incident beam direction. The X-rays were linearly polarized with the polarization vector parallel to the scattering plane. Photoemission data were collected under normal emission. To prevent radiation damage of the sample, we repeatedly changed the position of the beam on the surface. The sample was allowed to cool down below 100°C before collecting the photoemission spectra. The total instrumental energy resolution was set to 0.2 eV and 0.4 eV for photon energies of 650 eV and 915 eV, respectively. Sample annealing was performed using a manipulator with a PBN ceramic heater. Samples were annealed for 15 minutes per cycle. In addition to the synchrotron source, two laboratory XPS systems were used for XPS characterization. The first system features a non-monochromatic X-ray source with an Al anode (hν = 1486.4 eV), a SPECS Phoibos MDC 9 electron energy analyzer, configured for both normal and grazing detection angles, and a base pressure below $1\times10^{-9}$ mbar. The second system is equipped with a monochromatic X-ray source with Al kα, a hemispherical energy analyzer EA 15 UHV (PREVAC system) for normal detection angles, and a base pressure below $1\times10^{-9}$ mbar. In the case of monochromatic X-ray source, data were obtained with a pass energy of 100 eV. Apart from a wide survey spectrum, spectral regions of Co 2p, K 2p, Sr 3d, Ti 2p, Ta 4f, O 1s, and C 1s were acquired. The KolXPD software was used for data fitting after Shirley's background subtraction in all cases. The O 1s and Ti 2p components were fitted with Voigt functions, while the Sr 3d, K 2p, and Ta 4f spectra were fitted with Voigt doublets functions.

# 3. Results

## 3. 1. X-ray photoelectron spectroscopy

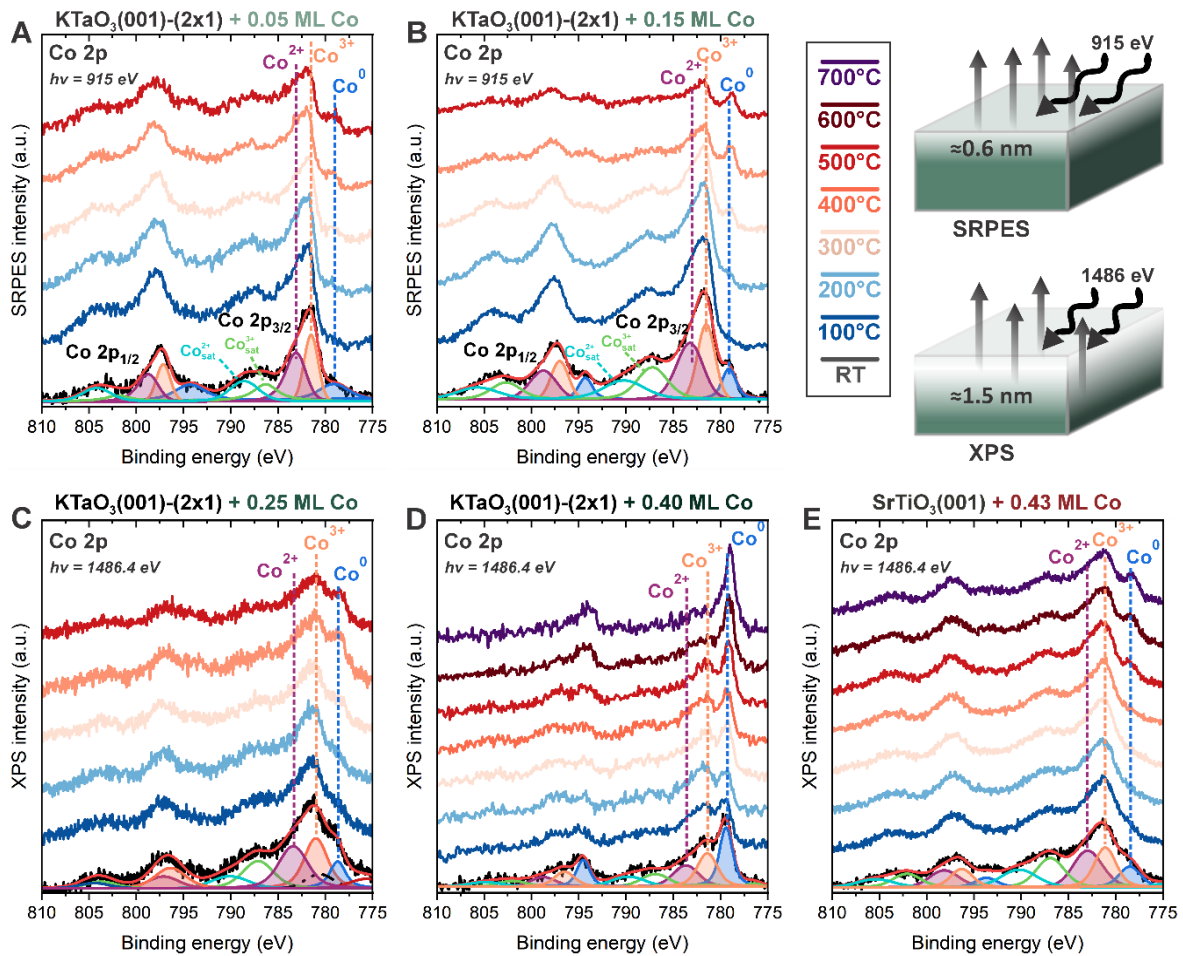

*Figure 1. Evolution of the cobalt Co 2p spectra upon deposition at RT, followed by cycles of annealing to gradually increasing temperatures. A-D) Hydroxylated KTaO$_3$(001)-(2×1) surface with 0.05, 0.15, 0.25 and 0.4 ML of cobalt, respectively, E) Wet-treated SrTiO$_3$(001) with 0.43 ML of Co. Panels A, B show data measured at a synchrotron (photon energy 915 eV), panels C, D, E were obtained using a laboratory x-ray source with an Al K$_α$ line (1486.4 eV). Schematic illustration indicats inelastic mean free path of SRPES and XPS, respectively.*

Chemical analysis of the deposited cobalt and its evolution on the perovskite surfaces is traced by X-ray photoelectron spectroscopy, employing both synchrotron radiation (SRPES) and laboratory XPS sources. The sample is annealed to gradually increasing temperatures in 100°C increments, allowing it to cool down after each step to perform the XPS measurement. **Figure 1** shows Co 2p photoelectron spectra for various cases: panels A-D were measured on the hydroxylated KTaO$_3$(001)-(2×1) surface with 0.05, 0.15, 0.25 and 0.4 ML of deposited cobalt respectively. Panel E was measured for 0.43 ML of cobalt deposited on the SrTiO$_3$(001) surface. Each panel shows results after deposition at room temperature and evolution after gradual annealing to higher temperatures.

All panels in Figure 1 show certain common features: the character of the deposited cobalt is predominantly ionic, with a mixture of 2+ and 3+ oxidation states. A metallic (0 oxidation state) contribution is present directly after deposition, and the associated peak is broad, reflecting the size distribution of the initial population of the formed nanoparticles. For higher coverages, the metallic contribution decreases after annealing to ~200°C and reappears after annealing to 300 to 500°C at a slightly lower binding energy (by ~0.5 eV). The small shift is attributed to size effect[57,58] associated with cobalt clustering. Note that the PES spectra in Figure 1A, B were recorded using synchrotron radiation with hν = 915 eV optimized for enhanced Co photoionization cross-section and surface-sensitive data acquisition (inelastic mean free path ≈ 0.6 nm), whereas in Figures 1C, D, E a laboratory monochromatic x-ray and a non-monochromatic x-ray with hν = 1486.4 eV was used (inelastic mean free path ≈ 1.5 nm).

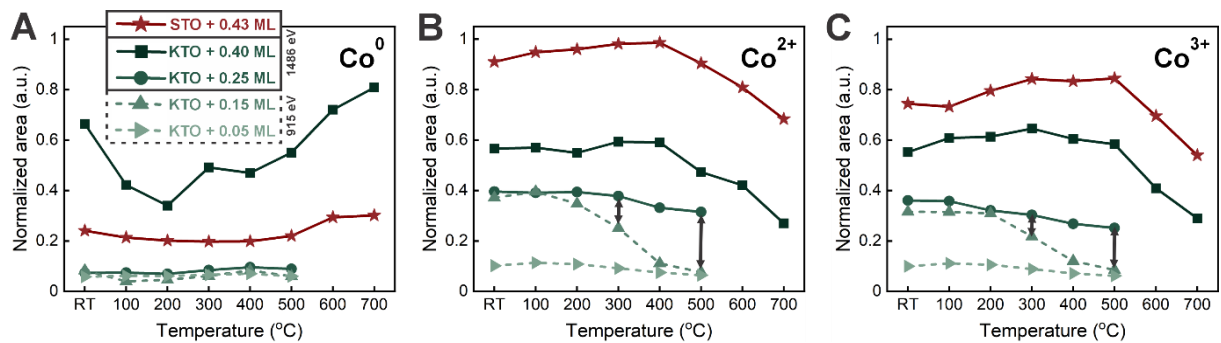

*Figure 2. The area of the Co 2p peak as a function of the annealing temperature.* A) Concentration of metallic $Co^0$, B) ionic $Co^{2+}$, and C) ionic $Co^{3+}$ species.

The quantitative evaluation of the metallic and ionic contributions of cobalt is summarized in **Figure 2**. Plotted is a normalized area of the respective peaks, where the normalization was performed with respect to the background signal to eliminate instrumental effects such as the beam intensity. For $KTaO_3$, the ionic contribution is dominant for all coverages up to 0.25 ML, a significant metallic peak develops only for the dataset with the highest coverage of 0.4 ML. Upon annealing, the concentration of the ionic species slightly decreases at temperatures below 500°C, followed by a more rapid decrease above this threshold. Interestingly, there is an increasing discrepancy between more surface-sensitive and deeper probing measurements. Arrows in panels B and C show that ionic contributions are much higher in the subsurface, indicating a segregation of cobalt in the first few layers below the surface. The metallic peak grows concurrently with the decrease of the ionic species. Elements of the substrate (K, Ta, Sr, Ti, O) do not show any significant changes in the shape of the core level peaks, see Figures S2 and S3; only the O 1s spectrum shows a certain level of OH-related component, mainly hydroxyls on the KTaO$_3$ surface, which changes with temperature. The role of water is linked with the polarity compensation mechanism, since the initial (2x1) reconstruction is hydroxylated.[32] For SrTiO$_3$, the ionic contributions initially grow, up to 400°C, and the trend reverts above 500°C, where the ionic contributions start to decrease, and the metallic peak grows.

## 3. 2. STM / ncAFM measurements

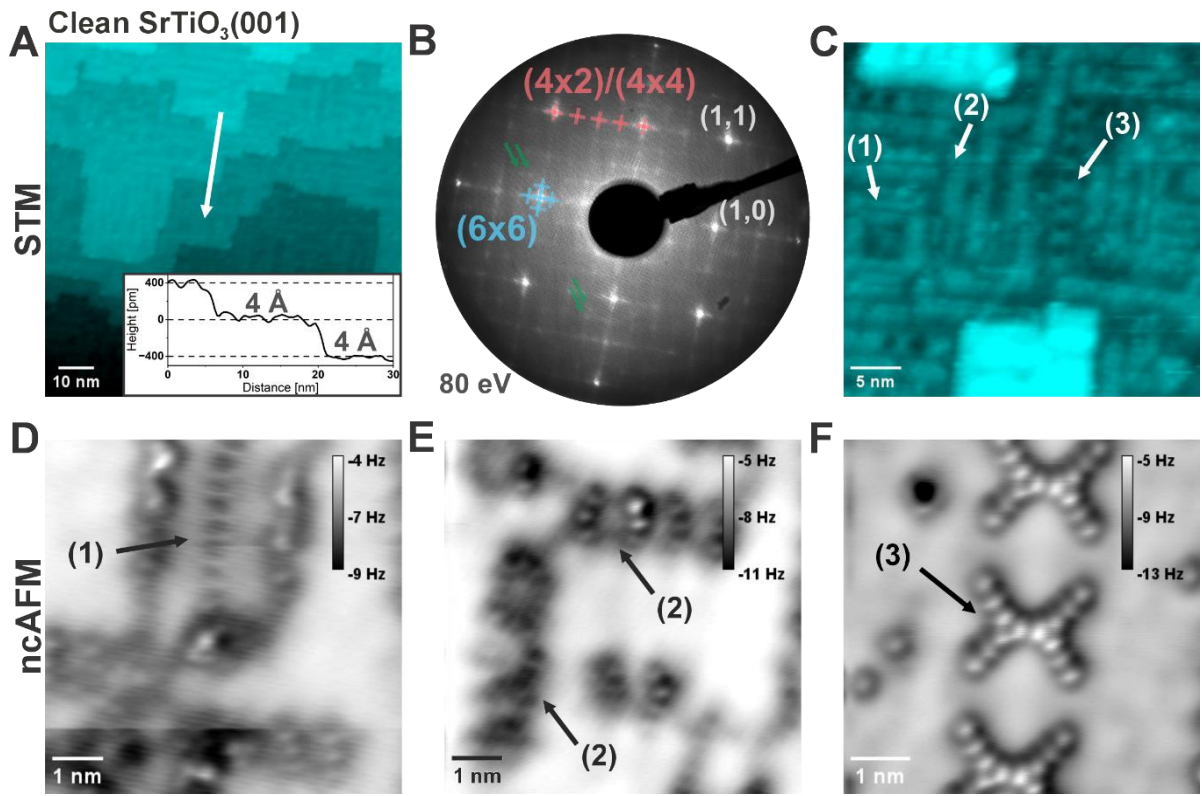

*Figure 3: Clean SrTiO$_3$(001) surface (before cobalt deposition)*. STM and ncAFM images of the SrTiO$_3$ (001) surface before the deposition of Co atoms. A) STM topography (U = 2.2 V, I = 2.5 pA) with a line profile in the inset. B) LEED image at 80eV confirming the presence of (2×n) and (4×n) reconstructions. C) Detailed STM image (U= 2 V, I = 5 pA) with 3 predominant types of reconstructions marked by arrows. D-F) Constant-height AFM images, showing atomic-scale details of the three reconstructions. Oscillation amplitude A=100 pm was used.

Atomic-scale details regarding the adsorption and dispersion of cobalt ions on perovskite surfaces, along with structural changes, were investigated by combined STM/AFM. First, the case of SrTiO$_3$ (**Figures 3-5**) will be discussed, followed by a comparison to KTaO$_3$ (Figure 6). Figure 3 illustrates STM and ncAFM images of the clean SrTiO$_3$(001) surface, as prepared by wet chemical etching followed by annealing in UHV in a partial $O_2$ pressure of $5.5\times10^{-7}$ mbar at 800°C. STM topography image (Figure 3A) reveals the terrace structure with a full-step height of 0.4 nm, confirming the consistency of our $SrTiO_3$(001) surface with previous studies.[59] The LEED pattern in Figure 3B suggests that the dominant periodic ordering at the surface is (2×n) and (4×n), also consistent with previous reports on annealed $SrTiO_3$(001) surfaces.[51-53]

A more detailed view of these reconstructions can be obtained from atomically resolved STM and ncAFM images, see Figure 3C-F, where a mixture of three dominant reconstructions was identified. In STM (Figure 3C), they appear as (1) lines with a width of a single unit cell (0.4 nm) and a periodicity of (4×1) or occasionally (3×1), (2) structures arranged in a (4×2) or (4×4) periodicity, and (3) linear arrangement of 'X'-shaped protrusions with a typical periodicity of (6×6) or (6×7). Panels D-F show atomic-scale details of these reconstructions, marked by the respective numbers.

The surface with deposited cobalt is shown in Figures 4 and 5. **Figure 4** shows an overview of STM topographic images of flat $SrTiO_3$(001) terraces, decorated by Co clusters of various sizes. **Figure 5** depicts a series of constant-height ncAFM images that focus on the atomic-scale details. The cobalt clusters observed in Figure 4A-D show the expected trend upon annealing: the cluster size increases, and their density decreases with annealing (Figure 4E). The plot in Figure 4E also indicates the average amount of cobalt atoms in a cluster, estimated as the deposited amount of Co divided by the cluster density. This is compared to the number of cobalt atoms in a cluster estimated from its volume, assuming the particle shape as a hemispherical cap using its height measured from the STM image.[60] We note that the cluster width is typically influenced by the tip convolution effects; more detailed analysis and other models are shown in Table ST1. The cluster statistics in Figure 4E shows that a large portion of the deposited cobalt stays incorporated in these clusters, excluding significant level of dissolution into the bulk in the investigated temperature range.

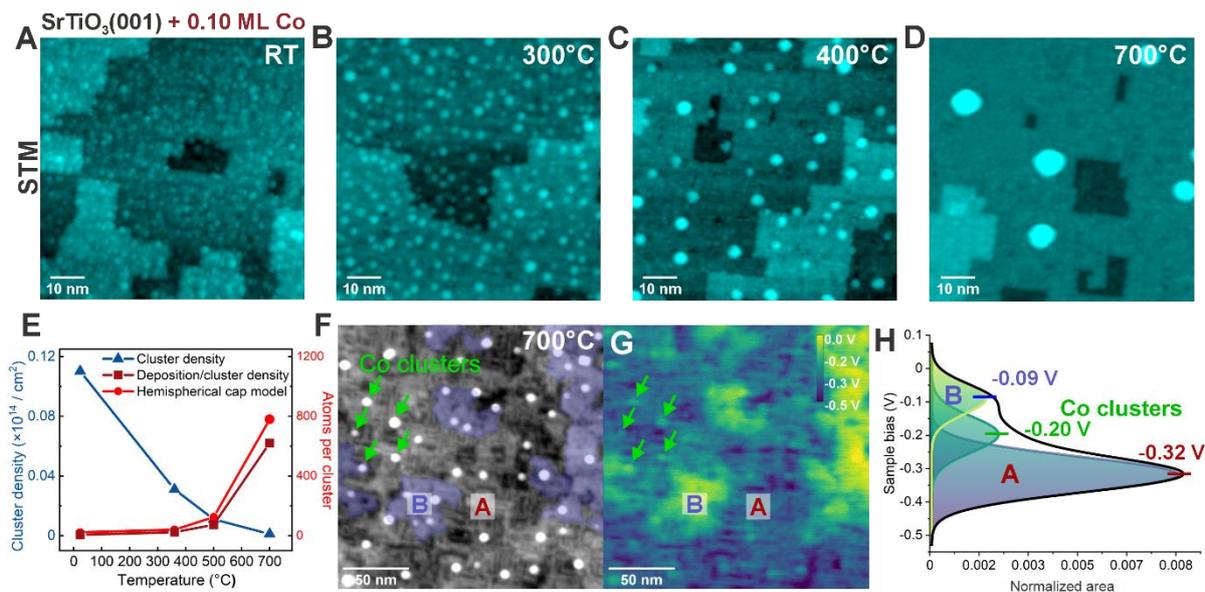

***Figure 4. Overview images of SrTiO₃(001) with 0.1 ML cobalt.*** *A-D) Topographic STM images of the surface after deposition at room temperature and annealing to 300, 400, and 700°C, respectively (75×75 nm²). E) Evolution of the cluster density with the annealing temperature, and of the corresponding average cluster size (number of Co atoms) calculated using various models. F), G) Topographic AFM image and a map of local contact potential difference (LCPD) of the surface annealed to 700°C, as measured by Kelvin probe microscopy. H) Histogram of the work function, the bottom A peak corresponds to a clean surface, the B shoulder is attributed to areas with cobalt and cobalt clusters are marked by arrows.*

An important question is whether the remaining surface between the clusters incorporates any single cobalt atoms to compensate its surface polarity. An interesting piece of information can be obtained from Kelvin Probe Force Microscopy (KPFM), see the Figure. 4F, G. After the last annealing step to 700°C, regions with two distinctly different values of work function form at the surface. The new surface phase has a local contact potential difference of +0.23V compared to the clean surface before deposition of the cobalt. The higher potential value charactrizes both the cobalt clusters and areas with the new phase, which is shown in more detail in Figure 5 D.

Atomic-scale details of the interaction of the deposited cobalt in the flat SrTiO₃(001) surface areas aside the large clusters are presented in Figure 5. Panel A shows the surface directly after deposition of 0.1 ML cobalt at room temperature. The surface is homogeneously covered by bright and dark protrusions of a single atom size; they cover the original reconstructions, which are now only visible as shadows of brighter and darker stripes. The images show many features of single atom size (marked by arrows), indicating that the cobalt atoms are kinetically frozen at the positions where they land on the surface. This indicates that the deposited atoms immediately stick at single sites, without any significant diffusion. Figure 5B shows that annealing to an intermediate temperature of 400°C results in a surface that partially recovers the original reconstructions, yet they are damaged at many locations and are surrounded by atom-sized features. Finally, annealing to 700°C results in the formation of two distinctly different regions at the surface, which were already outlined by the KPFM measurements in Figure 4F. Atomic-scale detail of the areas with a lower work function is shown in Figure 5C. These reconstructions (marked by arrows 1,2,3, consistently with Figure 4) are identical to the original surface before the cobalt deposition. Figure 5D shows a detail of an area with the higher work function. These areas are covered by a new surface phase that covers approximately 20% of the surface and was not present before the cobalt deposition and has not been previously reported in the literature. We, therefore, attribute this phase to a cobalt-stabilized structure. The presence of cobalt is further supported by the KPFM measurements, where the local work function of this new phase is similar to that of the large cobalt clusters. This phase has a structure that is not a trivial translation of a repeating surface unit cell, details are shown in the Figure. S4. Similar quasi-periodic or even quasi-crystalline behaviour has been reported for multiple surfaces of perovskites and doped perovskites.[61,62]

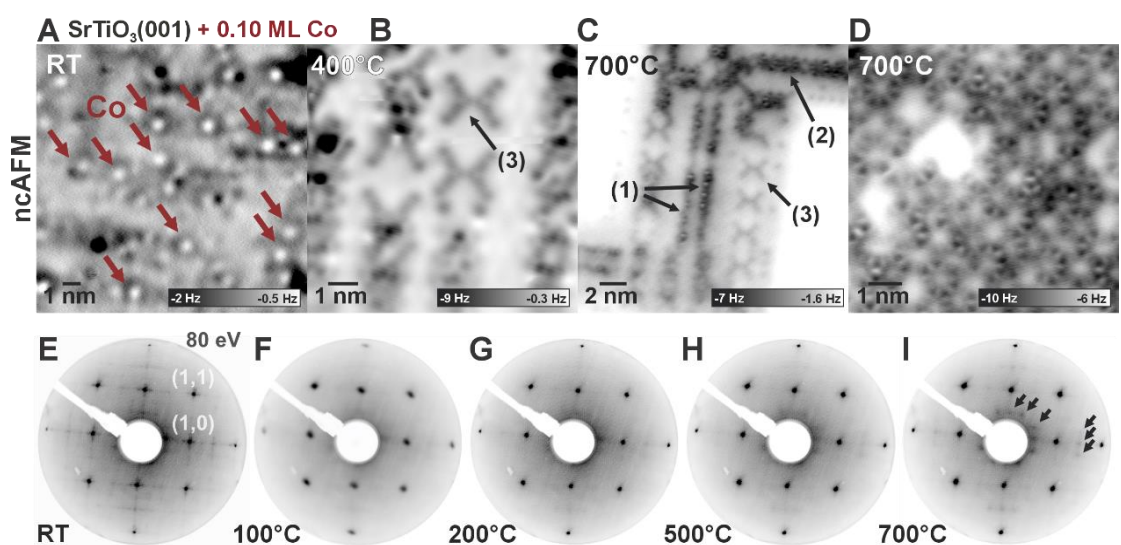

*Figure 5. Atomic-scale constant-height AFM images of the flat regions of SrTiO$_3$ with 0.1 ML cobalt. A) After deposition at room temperature. B) After annealing to 400°C. C), D) Different areas on a sample annealed to 700°C; panel C shows reconstructions identical to a clean surface, panel D shows a new reconstruction induced by the presence of cobalt. E-I) LEED images of the surface after cobalt deposition at room temperature and annealing to 100, 200, 500, and 700 °C (with new spots marked by arrows), respectively (beam energy of 80 eV).*

The local information obtained by AFM is complemented by LEED images, see Figure. 5E-I. The original LEED pattern obtained after deposition at room temperature is comparable with the pattern obtained on the clean surface (Figure 3B). Mild annealing to 100°C (Figure 5F) removes the fractional spots, which is consistent with the picture of a stronger cobalt interaction with the surface and its incorporation in the reconstructions, resulting in a destruction of long-range periodicity. The fractional spots gradually recover after annealing to higher temperatures, starting at ~500°C (Figure 5H), which is attributed to the extraction of the cobalt atoms from the surface and formation of clusters. There is an observable trace of c(4×2) phase,[63] after annealing to 700°C from the LEED pattern, see Figure 5I. This symmetry does not match the AFM images of the topmost layers (Fig. 5D). The AFM detection range is below the LEED information depth (5-10Å) at an energy of 80 eV, suggesting a possible restructuring of the subsurface layer. This picture is supported by some of our AFM images, see Fig. S5. Since the c(4×2) phase is not present in the LEED image (see Fig. 3B) prior to the deposition of cobalt, this could be again attributed to the role played by the cobalt atoms incorporated in the lattice to induce a stable (sub-)surface reconstruction.

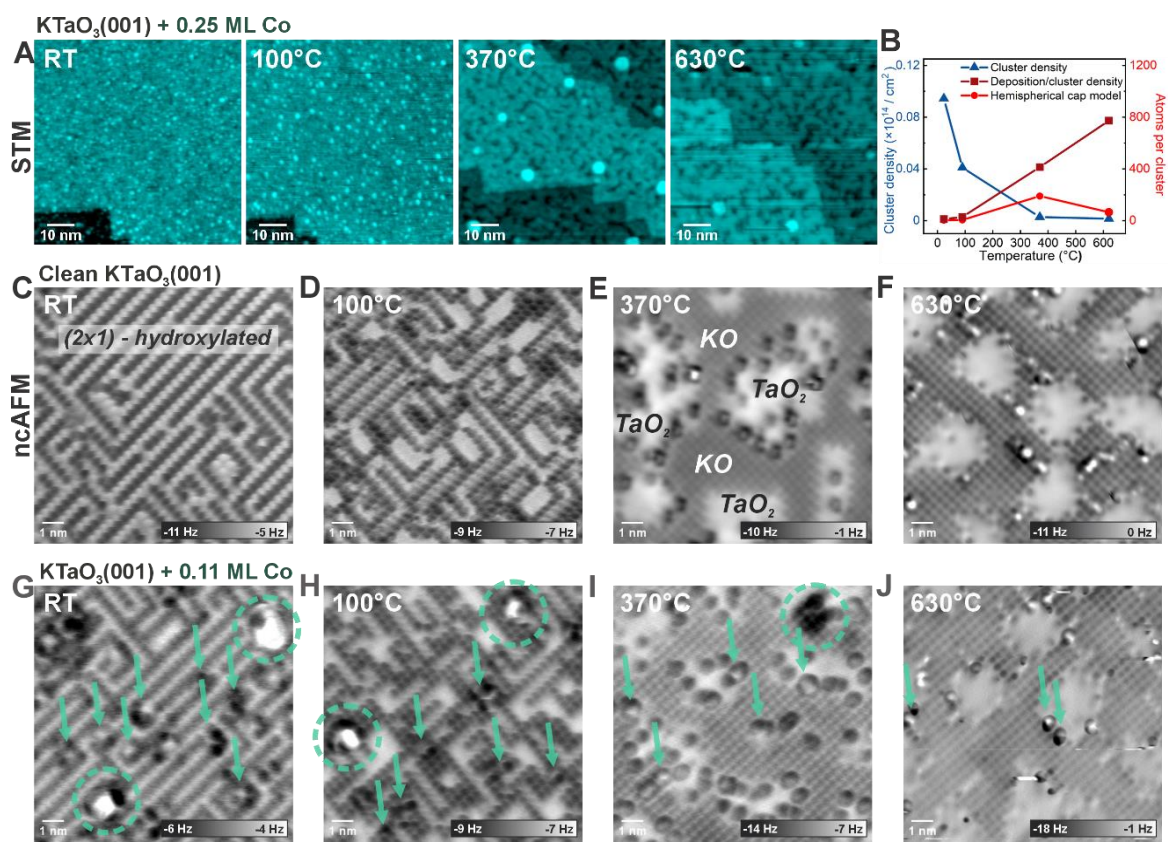

*Figure 6. Deposition of cobalt on the polarity-compensated KTaO$_3$(001) surface.* A) STM topography images of the surface after deposition of 0.25 ML at room temperature and annealing to 100, 370 and 630°C, respectively (75×75 nm$^2$). B) Evolution of the cluster density and average number of atoms per cluster with the annealing temperature using various models. C)-F) Atomic-scale constant-height AFM images of the polarity compensated KTaO$_3$(001)-(2x1) surface without cobalt formed at C) room temperature and after annealing to D) 100°C, E) 370°C and F) 630°C. G)-J) Atomic-scale constant-height AFM images of the the same surface after 0.11 ML cobalt deposition at G) room temperature and after annealing to H) 100°C, I) 370°C and J) 630°C.

For the case of KTaO$_3$(001) surface, cobalt deposition follows similar trends as for SrTiO$_3$(001), the major difference lies in the fact that STM/AFM does not directly show cobalt atoms (ions) incorporated in the surface layer at higher temperatures (**Figure 6**). After deposition, the cobalt atoms are homogeneously dispersed in the form of single atoms or small clusters (marked in green at Fig. 6 G, H). With annealing to higher temperatures (>300°C), water desorbs completely, the 2×1 reconstruction transforms into a 1×1 phase with KO and TaO$_2$ terminations, and the surface with cobalt appears similar to the clean surface at the atomic scale. From this point, it is difficult to unambiguously distinguish or identify any possible cobalt atoms from the potassium and tantalum.

Analysis of the cluster sizes (Figure 6B), however, indicates that the number of atoms incorporated in the clusters is lower than the deposited amount. This becomes pronounced especially after annealing to higher temperatures (above 600°C), indicating that a certain amount of cobalt might be distributed in several substrate layers in the near-surface region. Previous work indicated that even a clean KTaO$_3$(001) surface compensates its polarity through rearrangement of this near-surface region, including the formation of subsurface potassium vacancies.[14] It is likely that cobalt atoms enters this process and accommodates in a few topmost perovskite layers. This picture is consistent with the XPS data in Figures 1 and 2, where the amount of cobalt measured at surface-sensitive conditions decreases after annealing, while it remains almost constant for higher probing depths.

## 4. Discussion and conclusions

In this study, we have examined the impact of extrinsic metal atoms on the polarity compensation mechanisms operating at surface polar perovskite surfaces. After deposition at room temperature, cobalt forms a mix of single atoms and smaller clusters on both KTaO$_3$(001) and SrTiO$_3$(001) surfaces. For weakly polar SrTiO$_3$, increasing the temperature promotes both clustering and incorporation into the surface layer, via the formation of new reconstructed phases that coexist with the original reconstructions. The newly formed surface reconstruction exhibits a lower work function compared to the clean surface by 0.23 eV. Such a difference is interesting for catalysis and charge separation in photocatalysis: perovskites such as NaTaO$_3$, KTaO$_3$ or SrTiO$_3$ doped by various metals[4,62,65] achieve record photocatalytic efficiencies after annealing to temperatures above 700°C.

The polar KTaO$_3$(001) surface evolves in a similar way to the SrTiO$_3$(001), with the exception that atomically resolved microscopies do not show direct evidence for atomically dispersed Co atoms after annealing, and the surface retain its original (1×1) periodicity. Yet, indirect evidence based on a combination of statistical processing and XPS analysis indicates the cobalt does incorporate in several perovskite layers near the surface to compensate the surface polarity, which stands in agreement with the findings of the clean annealed KTaO$_3$(001) surfaces[14].

The different behaviour of KTaO$_3$ and SrTiO$_3$ can be linked with the properties of the B-site cation in these perovskites: titanium (Ti$^{4+}$) is reducible to Ti$^{3+}$ and can change the bonding configuration from octahedral (present in bulk perovskites) to tetrahedral, which is a typical building block of surface reconstructions.[66-68] Contrarily, tantalum has a strong preference for the octahedral configuration and is not easily reducible by annealing in UHV.[69] Consequently, SrTiO$_3$ surfaces have an easier path to reconstruct and incorporate the extrinsic cobalt, while KTaO$_3$ keeps the bulk-terminated surface structure and presumably favours polarity compensation through cation exchange in the subsurface region. The atomic-scale picture provided in this study represents a missing lead in understanding the behaviour of model perovskites in applications related to photocatalysis or electrocatalysis towards the single-atom catalytic systems, which help to reduce dependence on scarce and high-cost materials.

**Acknowledgements**

The work was funded by the Czech Science Foundation, project GACR 20-21727X, and by Czech Ministry of Education, Youth and Sports, project Quantum materials for applications in sustainable technologies (QM4ST), project no. CZ.02.01.01/00/22_008/0004572 by Programme Johannes Amos Comenius, call Excellent Research. A.A. and P.S.K acknowledges the support from the Grant Agency of Charles University, project GAUK 326122. This research was funded in part by the Polish National Science Centre, project SONATA 2022/47/D/ST5/02439. We acknowledge Elettra and Sincrotrone Trieste for providing access to its synchrotron radiation facilities and for financial support by the IUS (P2022003) project. The authors thank Lynn A. Boatner for providing KTaO$_3$ single crystals.

**Data Availability Statement**

Data is available at the open repositories: Zenodo (https://doi.org/10.5281/zenodo.18108562) and RODBUK (https://doi.org/10.57903/UJ/TMALVJ).

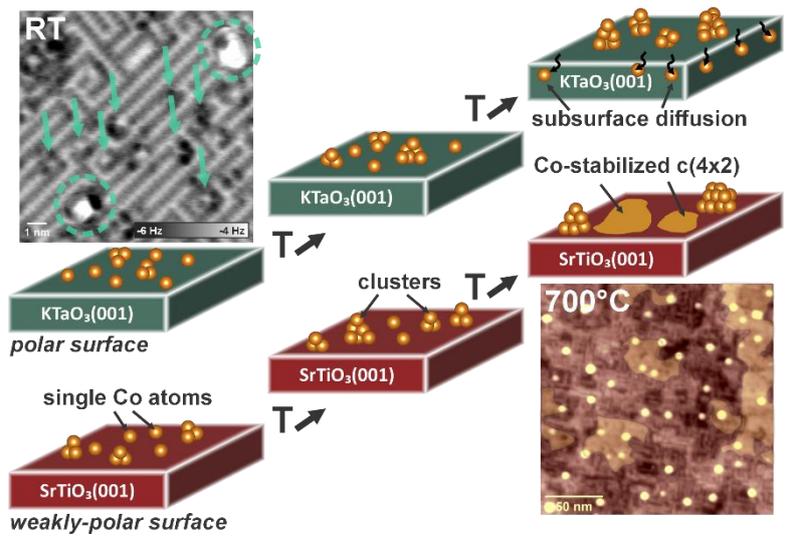

Supporting Information

# Competition between clustering and dispersion of cobalt atoms on perovskite surfaces: SrTiO$_3$(001) and KTaO$_3$(001)

*Aji Alexander[1], Pankaj Kumar Samal[1], Llorenç Albons[1], Jesús Redondo[1,2], Jan Škvára[1], Igor Píš[3,4], Lukáš Fusek[1], Josef Mysliveček[1], Viktor Johánek[1], Dominik Wrana[1,5]\*, Martin Setvin[1]*

[1] Department of Surface and Plasma Science, Charles University, 18000 Prague, Czech Republic

[2] Department of Polymers and Advanced Materials, Facultad de Química, University of the Basque Country UPV/EHU, 20018 San Sebastián, Spain

[3] CNR – Istituto Officina dei Materiali (IOM), 34149 Trieste, Italy

[4] Institute of Electrical Engineering, Slovak Academy of Sciences, 841 04 Bratislava, Slovakia

[5] Marian Smoluchowski Institute of Physics, Jagiellonian University, 30-348 Kraków, Poland

\* dominik.wrana@uj.edu.pl

**Supplementary Figures and Table:**

Figure S1: XPS spectra of SrTiO$_3$ (001) surface after wet chemistry and annealing in O$_2$

Figure S2: XPS of 0.25ML Co on KTaO$_3$ (001) surface annealed up to 500°C

Figure S3: The evolution of O 1s components

Figure S4: Quasi-periodic nature of the Cobalt-induced reconstruction on SrTiO$_3$(001)

Figure S5: Cobalt-induced c-(4×2) reconstruction on SrTiO$_3$(001)

Table ST1: Statistical analysis of the cobalt cluster sizes on the perovskite surface (SrTiO$_3$ and KTaO$_3$)

**XPS of the SrTiO₃(001) surface after wet chemistry and annealing in O₂**

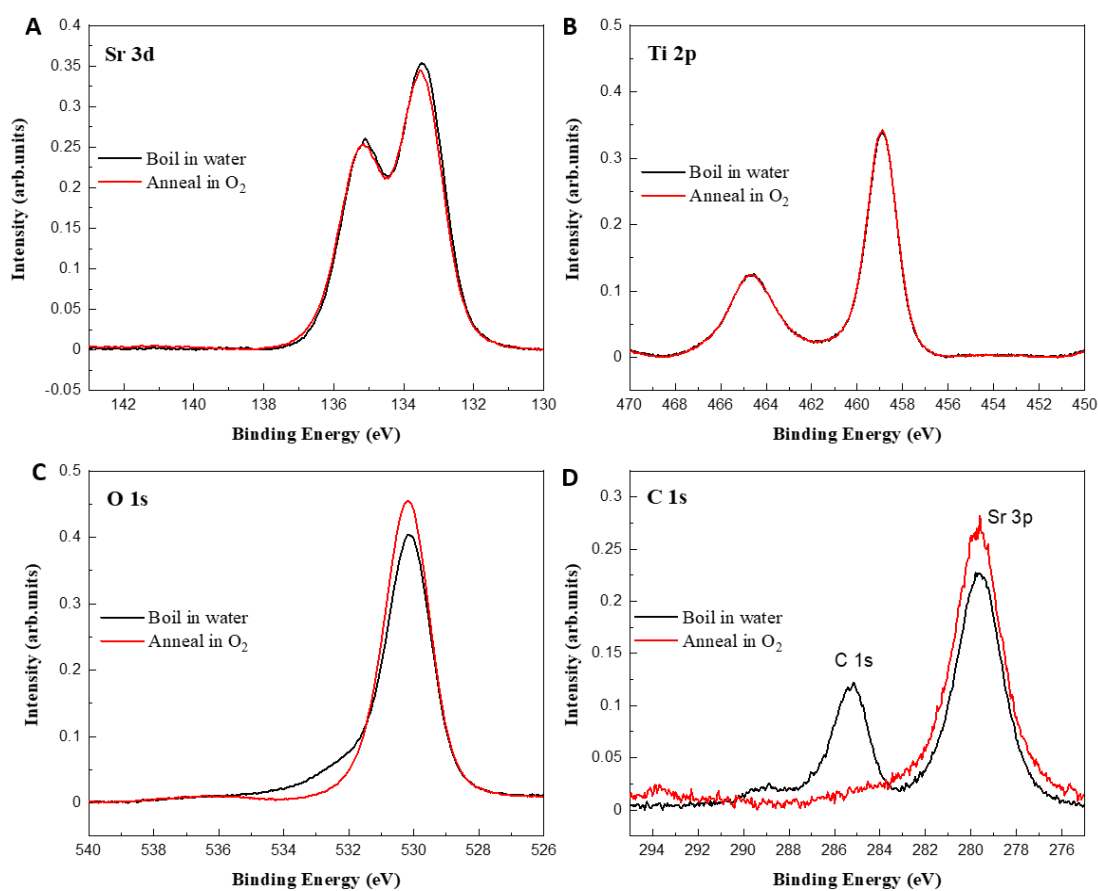

*Figure S1.* ***XPS spectra of SrTiO₃ (001) surface after wet chemistry and annealing in O₂.** A) Sr 3d B) Ti 2p C) O 1s and D) C 1s*

**Figure S1** shows XPS spectra of the SrTiO$_3$(001) surface directly after the wet chemical etching in boiling water, and after annealing to 800°C in 5×10$^{-7}$ mbar O$_2$. The changes in the Sr 3d peak (panel A) and Ti 2p (panel B) are negligible. In the O1s peak, annealing causes removal of the shoulder at the high-binding side, which is attributed to the removal of OH components and oxygen included in any carbon contamination (panel C). Most importantly, the annealing removes adventitious carbon (panel D).

## All XPS cationic components (Sr 3d, Ti 2p, K 2p, Ta 4f)

The strontium 3d spectrum has a doublet Voigt function centred at 133.6 eV. The Ti 2p main peak is deconvoluted into two singlet Voigt functions at 459.05 eV and 464.7 eV, with a splitting of 5.7 eV (**Figure S2** A, B). This indicates the presence of $Ti^{4+}$ oxide as a major component. Annealing in $O_2$ before deposition prevents chemical reduction of the sample, and thus no visible traces of $Ti^{2+}$ and $Ti^{3+}$.

In the case of the potassium (K 2p) component, two Voigt functions were centered and fitted at 292.5 eV and a higher binding energy at 294.0 eV. The tantalum, Ta 4f, spectra perfectly fit at 26.8 eV, which represents the bonding of $Ta^{5+}$ states throughout the experiment (Figure S2C,D). The oxygen 1s spectrum is shown in **Figure S3**. The one at RT is deconvoluted into two primary peaks at 530.2 eV and 531.0 eV (dotted line). The singlet Voigt function is used to fit both peaks. The one with the higher binding energy (eV) represents the surface oxygen forming coordination with the cations, whereas the lower one indicates lattice oxygen.

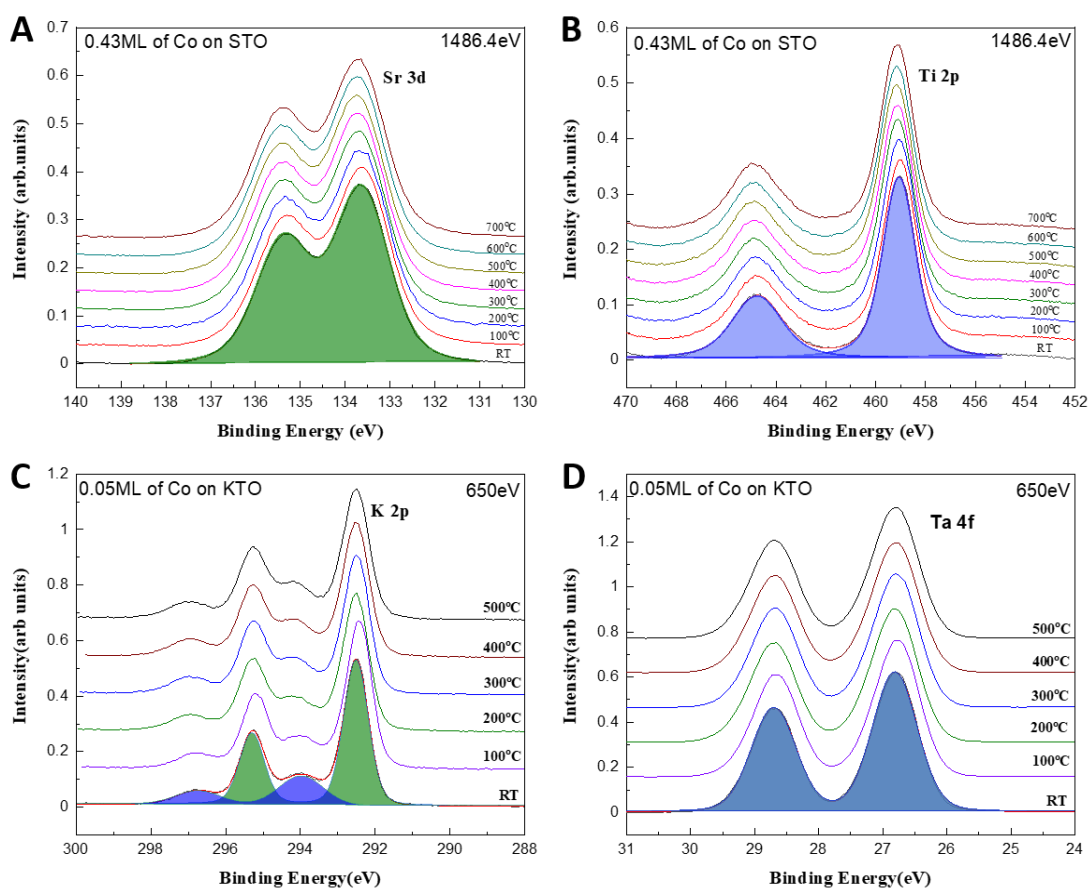

*Figure S2. **The evolution of other components with annealing**, A) Sr 3d and B) Ti 2p, on the reconstructed $SrTiO_3(001)$ surface. C) K 2p and D) Ta 4f on polarity compensated $KTaO_3(001)$ surface under different annealing conditions.*

## Details of Cobalt XPS fitting

The cobalt components were fitted with Voigt functions including the main and satellite peaks (see **Figure 1** in the main text). The ionic peaks are assigned at higher binding energies, whereas the metallic peak, $Co^0$, is fitted at 779.1 eV. The $Co^{3+}$ peak is fitted at 781.5 eV, whereas the $Co^{2+}$ peak is fitted at 783.0 eV. Both ionic components (+2 and +3) are coupled with the shake-up satellites at higher binding energies of 787.2 eV and 790.2 eV. For the cases of higher cobalt coverages (See Fig. 1D and E), the metallic peak is fitted at 780 eV, and the ionic peak is fitted at 781 eV ($Co^{3+}$) and 783.3 eV ($Co^{2+}$) for 0.25 ML. For 0.43 ML of cobalt on $SrTiO_3$, the metallic peak is varied between 778.5 and 778.3 eV, the ionic peaks are fitted at 782.6 eV ($Co^{3+}$) and 781.1 eV ($Co^{2+}$). Figure 1C represents XPS obtained using monochromatic Al Kα x-rays with a pass energy of 100 eV. Here, the metallic component is fitted at 782.1 eV, $Co^{2+}$ is fitted at 784.2 eV, and $Co^{3+}$ is fitted at 782.7 eV, respectively.

## Oxygen XPS spectra

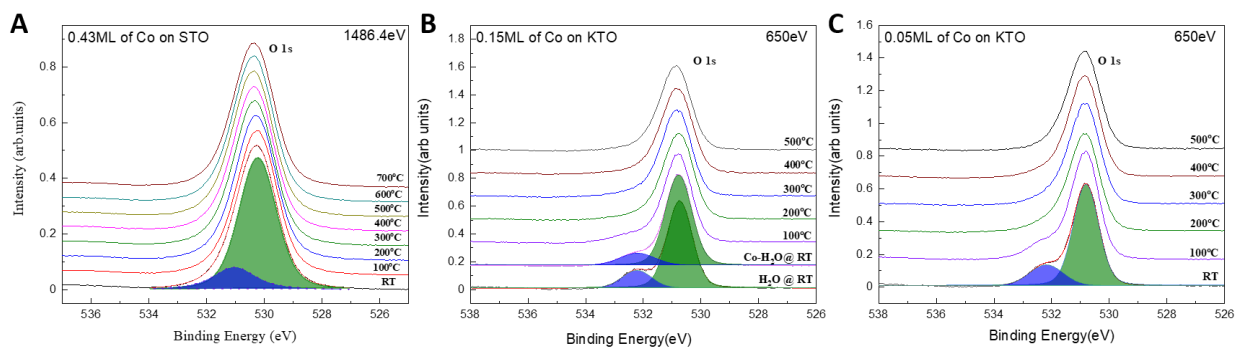

*Figure S3. **The evolution of O 1s components** in A) 0.43ML of Co on the reconstructed $TiO_2$ termination of $SrTiO_3(001)$ surface. B) 0.15 ML and C) 0.05 ML of Co deposited on polarity compensated $KTaO_3(001)$ surface under different annealing conditions.*

The oxygen O 1s spectra are fitted with two peaks, see Figure S3A. In the case of $SrTiO_3$ (panel A), the main peak at 530.5 eV is attributed to the bulk lattice oxygen and the smaller peak at higher binding energy is attributed to surface oxygen anions. For $KTaO_3$, the main peak at 530.5 eV is again attributed to the bulk lattice oxygen, while the smaller peak at higher binding energy is attributed to the OH contribution, which takes part in the water-induced (2×1) reconstruction[1] (panels B and C). The oxygen anion coordination appears stable in the case of the $SrTiO_3(001)$ surface. When comparing the OH contribution with and without Co in the case of higher coverage on $KTaO_3$ (panel B), it is evident that the area under the OH peak with 0.15ML of Co is smaller than the water-dosed system, indicating that the cobalt atoms are positioned on the $K(OH)_2$ water-reconstructed surface, thereby screening the signals from the water. Annealing the KTaO$_3$ surface results in gradual desorption of the water in the temperature range from 100 to 200°C, leading to disappearance of the OH contribution[2].

## STM image analysis: cobalt clusters

| Co Deposition (calibrated from QCM) | Temperature (Experiment Condition) | Total number of metal atoms From QCM ($\times 10^{14}$ atoms/cm$^2$) | Particle Number Density From STM ($\times 10^{14}$ particles/cm$^2$) | Number of Metal atoms per Particle From QCM/STM (atoms/particle) | Mean Particle Diameter From STM (nm) | Mean particle Height From STM (nm) | Metal atoms per particle From STM (hemispherical-cap) (atoms/particle) $V=2/3\pi r^3$ (using mean particle height as the radius of hemispherical cap) | Metal atoms per particle From STM (spherical-cap) (atoms/particle) $V=1/6\pi rh(3a^2+h^2)$ Where a is the base of the cap and h is the height | Total number of metal atoms From STM ($\times 10^{14}$ atoms/cm$^2$) [particle density × number of metal atoms/particle] (spherical cap) | Total number of metals surfaces atoms on the sample From STM for particle ($\times 10^{14}$ atoms) [no of metal surface atoms × particle density × area of sample (0.56cm2 for STO and 0.24cm2 for KTO)] (Spherical cap) |
|---|---|---|---|---|---|---|---|---|---|---|
| **0.1ML on SrTiO$_3$** | RT (23°C) | 0.621 | 0.1102 | 5.6 | 2 | 0.5 | 23 | 77 | 8.5 | 4.76 |
| | 360°C | 0.621 | 0.031 | 23.8 | 3-4 | 0.6 | 41 | 202 | 6.262 | 3.51 |
| | 500°C | 0.621 | 0.011 | 73.9 | 4.5 | 0.87 | 125 | 660 | 7.2 | 4.03 |
| | 700°C | 0.621 | 0.0012 | 621 | 11 | 1.86 | 1224 | 8337 | 10 | 5.6 |
| **0.25ML on KTaO$_3$** | RT (23°C) | 1.68 | 0.0944 | 17.8 | 2 | 0.26 | 3.3 | 37 | 3.5 | 0.84 |
| | 90°C | 1.68 | 0.041 | 40.97 | 3 | 0.3 | 5.1 | 97 | 3.97 | 0.95 |
| | 370°C | 1.68 | 0.0028 | 600 | 8 | 1 | 190 | 1006 | 2.82 | 0.68 |
| | 620°C | 1.68 | 0.00093 | 1806.5 | 4.5 | 0.65 | 52 | 482 | 0.448 | 0.11 |

*Table ST1. **Statistical analysis of the cobalt cluster sizes on the perovskite surface (SrTiO3 and KTaO3).** The total number of metal atoms deposited is compared to the total number of surface metal atoms observed on the surface, assuming a hemispherical and spherical cap model for the particle/cluster shape.*

The average number of Co atoms per particle can be calculated from the Co amount deposited (as measured by the QCM) and the Co particle number density deduced from STM. This gives 6 atoms/particle (KTaO$_3$ 0.1ML), and 17 atoms/particle (SrTiO$_3$ 0.25ML) after deposition at room temperature, see **Table ST1**. The number of cobalt atoms per cluster can be also estimated from the diameter and height of the particles, as measured by STM. Here we considered two models for the particle shape: Hemispherical model and a spherical cap model. For the hemispherical model, the cluster height was considered as a radius and the cluster volume was evaluated as for the corresponding hemisphere. For the spherical cap model, the height and the diameter were estimated from the STM image and the cluster volume was evaluated according to the equation in Table ST1.

This type of analysis is only for orientation because of the tip convolution effect, which tends to broaden the island. The mean height of the cluster is more accurate, but still there is certain uncertainty caused by the fact that the clusters are predominantly metallic, while the substrate is a wide-gap semiconductor, affecting their relative apparent heights in STM. We consider the hemispherical cap mode as more relevant from these reasons.

**Cobalt-induced c-(4×2) reconstruction on SrTiO$_3$**

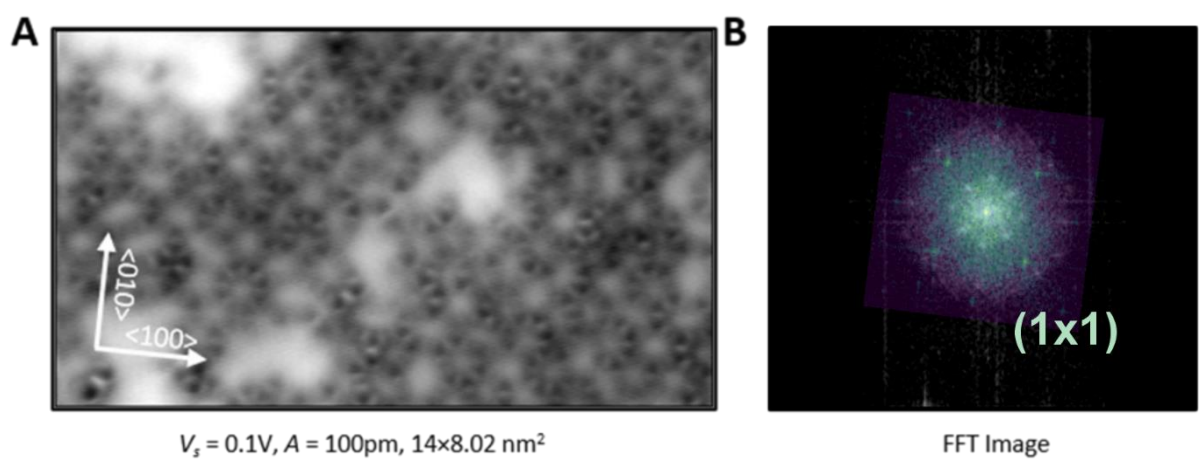

*Figure S4.* ***Quasi-periodic nature of the cobalt-induced reconstruction on SrTiO$_3$.*** *A) New reconstruction induced by the presence of cobalt when annealing the SrTiO3 (001) surface at 700°C. B) Fast Fourier Transform of the image.*

The SrTiO$_3$(001) surface after cobalt deposition depicted in **Figure S4** shows different periodicity in LEED images compared to the clean surface. After cobalt deposition, fractional (4x4)/(4x2) LEED spots disappear as soon as the sample is annealed to 100°C (see Fig. 5), but a new reconstruction starts to form when annealed above 300°C; pronounced spots in the diffraction pattern indicate a c(4x2) symmetry. Additional data are shown in **Figure S5**: LEED patterns in panels A and C show a comparison between the clean surface and a surface with deposited cobalt and subsequently annealed. The surface with cobalt shows a pronounced pattern with a c(4x2) periodicity. Blue and yellow grid superimposed on an image represents the centered reconstruction with cell dimensions of 1.64 nm and 0.83 nm respectively, which fits the c(4x2) surface, which is created for annealed SrTiO$_3$(001)[3] and was previously attributed to the excess of TiO$_2$ compared to SrO at the surface.[4] Here we observe that this transition is promoted by the presence of cobalt. While the STM/AFM images of the surface layer presented in the main text (Fig 4) do not show any clear presence of the c(4×2) reconstruction, it is possible to find indications of the c(4×2) ordering in the first subsurface layer. Figure S5D shows an area between the topmost surface reconstructions, where the c(4×2) pattern is visible. Notably, the same pattern was not observed on the clean surface, see Fig. S5B.

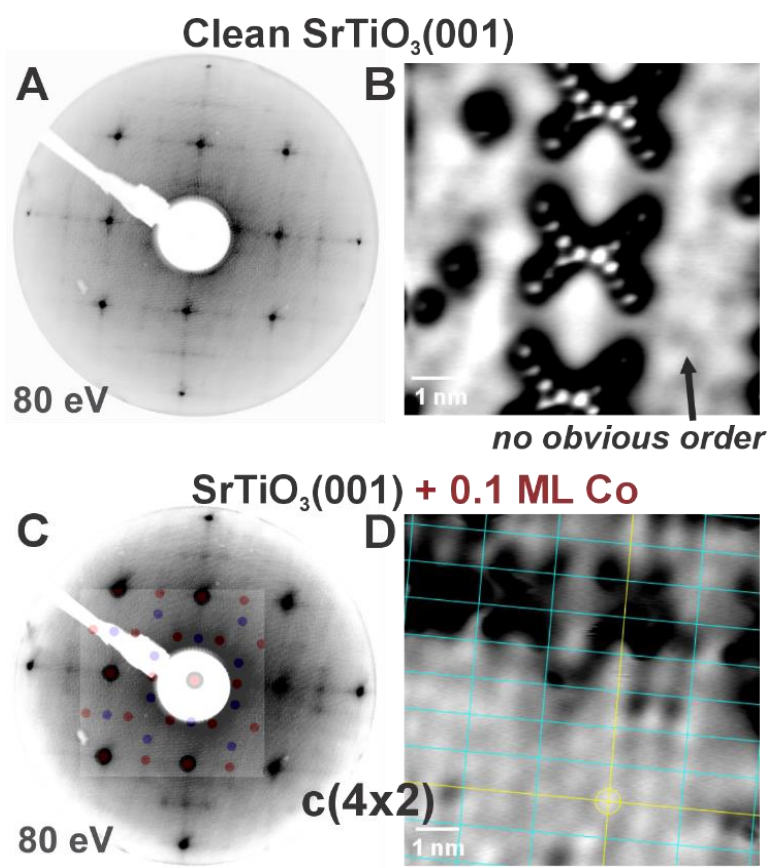

*Figure S5. Comparison between the same SrTiO$_3$(001) as prepared vs after deposition of 0.1 ML of cobalt.* A) LEED pattern showing predominantly (1x1) and (2xn)/(4xn) periodicity. B) Atomic-scale constant-height ncAFM (8x8 nm$^2$) with X-shaped adatom features but no obvious order in between. C) LEED pattern of the same surface with 0.1 ML of cobalt with c(4x2) reconstruction (two equivalent domains) with respective simulated pattern superimposed. D) ncAFM image (8x8 nm$^2$) of the SrTiO$_3$(001) with 0.1 ML cobalt, annealed to 300°C. The grid represents a c(4x2) reconstruction of the surface, highlighting the positions of atoms.

**References (Supplementary Information only)**